\input epsf
\documentstyle[prl,floats,aps]{revtex}
\begin{document}
\title{A rigorous solution of the quantum Einstein equations}

\author{Rodolfo Gambini\\{\em Instituto de F\'{\i}sica, Facultad de
Ciencias,\\Tristan Narvaja 1674, Montevideo, Uruguay}}

\author{Jorge Pullin\\  {\em Center for Gravitational Physics and Geometry}\\
{\em Department of Physics, 104 Davey Lab,}\\ {\em The Pennsylvania State
University,}\\
{\em University Park, PA 16802}}

\maketitle
\begin{abstract}
We show that the second coefficient of the Conway knot polynomial is
annihilated by the Hamiltonian constraint of canonically quantized
general relativity in the loop representation. The calculations are
carried out in a fully regularized lattice framework. Crucial to the
calculation is the explicit form of the skein relations of the second
coefficient, which relate it to the Gauss linking number.
Contrary to the lengthy formal continuum
calculation, the rigorous
lattice version can be summarized in a few pictures.
\end{abstract}

\vspace{-9cm}
\begin{flushright}
\baselineskip=15pt
CGPG-95/11-2  \\
gr-qc/9511042\\
\end{flushright}
\vspace{8cm}

The introduction of the Ashtekar variables \cite{As} and the loop
representation have opened new perspectives of finding exact quantum states
of the gravitational field.  In the loop representation, the
diffeomorphism constraint implies that wavefunctions have to be knot
invariants \cite{RoSm}.  A few years ago it was noticed
\cite{BrGaPuprl} that $a_2$, the second coefficient of a particular
knot polynomial ---the Conway polynomial--- was formally annihilated
by the Hamiltonian constraint of quantum gravity in the loop
representation.  The calculations were unregularized and involved
divergent factors, so the result was really of the form ``zero times
infinity''.  Later, another set of formal calculations \cite{GaPuba}
showed that this
solution was a reflection of the fact that in the quantum theory
formulated in terms of the Ashtekar connection, the exponential of the
Chern-Simons form was formally annihilated by all the constraints of
quantum gravity with a cosmological constant.  The expression of this
fact in the loop representation is that the Kauffman bracket knot
polynomial should be a solution of all the constraints with
cosmological constant.  While checking this fact \cite{BrGaPuessay},
again at a formal
level, it was found that $a_2$, the second coefficient of the Conway
polynomial, had to be annihilated by the Hamiltonian constraint
with zero cosmological constant.  This result was later confirmed in a
regularized setting via the extended loop representation
\cite{DiGaGr}, but it required lengthy manipulations and the
introduction of a counterterm in the Hamiltonian constraint. As a
consequence of all this the second coefficient of the Conway polynomial
appeared as one of the first very nontrivial exact states of quantum
gravity in the loop representation and hinted towards a deep connection
between notions of knot theory and quantum gravity, this time at a
dynamical level.

In this paper we will show that $a_2$ is rigorously annihilated by the
Hamiltonian constraint of quantum gravity in a fully regularized lattice
formulation of the theory.  In the lattice theory, which is described in
detail in \cite{FoGaPu}, the Hamiltonian constraint has a simple
geometric action.  As in the continuum, it only has a non-trivial action
at points where the loops intersect, and the effect is to produce two
terms per each pair of independent tangent vectors entering the
intersection: one in which the wavefunction is evaluated at a loop in
which one of the lines of the intersection has been shifted forward
along two links of the lattice in the direction of the other tangent at
the intersection, and one of the two sub-loops determined by the
intersection is re-routed; the other term is similar, but the loop is
shifted in the opposite direction.  The total action of the Hamiltonian
is given by the difference of both terms added over all possible pairs
of tangents entering the intersection (in the lattice at most three
pairs for intersections in which the loops traverse ``straight through'').
The action is shown in figure 1 for a double
self-intersection and in figure 4 for a triple self-intersection.
It is easy to check
that if one takes  the continuum limit by leaving loops fixed and
refining
the lattice they live in, the Hamiltonian has as leading contribution
the usual Hamiltonian constraint in terms of Ashtekar new variables.  In
order to do it, one evaluates the action of the Hamiltonian on a
holonomy of a connection $A$ on the lattice and shows that it reduces to
the usual formal action  of the Hamiltonian constraint in the connection
representation $\hat{H} = \epsilon_{ijk} F_{ab}^i \hat{E}^a_j
\hat{E}^b_k$ \cite{FoGaPu} acting on a holonomy.

\begin{figure}[t]
${}^{}$\hspace{4cm}\epsfxsize=250pt \epsfbox{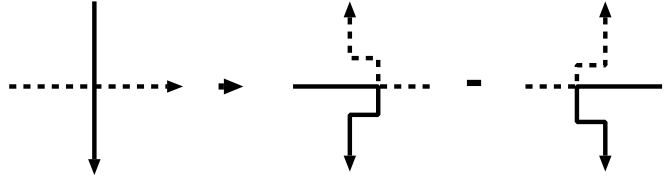}
\caption{The action of the Hamiltonian constraint on a double intersection,
which is equivalent to a deformation of the loop which can be obtained through
a diffeomorphism.}
\end{figure}

One can also
introduce a diffeomorphism constraint on the lattice, and show that it
satisfies the correct diffeomorphism algebra in the continuum limit, but
we will not discuss it here for reasons of space. Details are given in
\cite{FoGaPu}.

An important property of the Hamiltonian constraint is already
apparent from figure 1: the action of the constraint at points with
double intersections in the loop produces two contributions that are
deformable to each other, if the state one is considering is
diffeomorphism invariant in the continuum. Therefore the Hamiltonian
automatically vanishes at double intersections.
This fact has a counterpart in the continuum. If one
looks at the construction of quantum state in terms of the transform
of the Chern-Simons form, all the coefficients of the Jones polynomial
were automatically annihilated on double intersections, since the term
involving the cosmological constant is proportional to a determinant
of the metric, which vanishes on intersections less than triple
\cite{BrGaPuessay}.

\begin{figure}[b]
${}^{}$\hspace{3cm}\epsfxsize=250pt \epsfbox{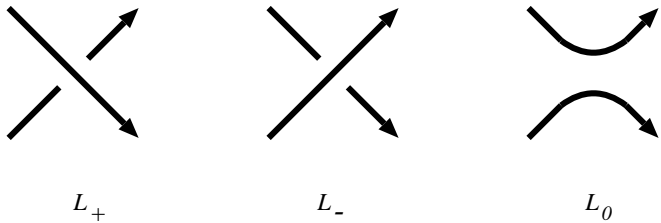}
\caption{The elements $L_{\pm}$ and $L_0$ that intervene in the
skein relation of the Conway polynomial without intersections.}
\end{figure}

Let us now discuss the candidate for solution. The Conway polynomial in
the variable $z$ is
defined by the following skein relation \cite{Kau},
\begin{equation}
C(L_+) - C(L_-) = z C(L_0).
\end{equation}

This relation is to be read in the following way: given a concrete
knot take a planar projection and focus at a given crossing in the
knot diagram; the value of the polynomial when the crossing is
replaced by an $L_+$ minus the value of the polynomial when the
crossing is replaced by $L_-$ is equal to the value of the polynomial
where the crossing is replaced by an $L_0$.  The definition of the
$L$'s is given in figure 2.  This relation, together with the
normalization condition that the polynomial is equal to one for the
unknot completely determines the polynomial for any knot without
intersections. In order to consider the polynomial as a state of
quantum gravity we have to define its value for loops with
intersections. For the calculations at hand we will only need
explicitly its definition for double intersections. There are many
possible extensions of a given polynomial to intersecting loops. In
principle, studying the transform of the Chern-Simons state could
provide the appropriate extensions of interest for quantum gravity. At
present however, the only available calculations of the transform for
intersecting loops are first order perturbation theory ones
\cite{BrGaPu}. These calculations are compatible with the following
skein relations,
\begin{eqnarray}
C(L_I)&=&{1 \over 2} (C(L_+)+C(L_+))\\
C(L_W)&=&C(L_0)
\end{eqnarray}
where the elements $L_I$ and $L_W$ are shown in figure 3 and correspond
to a ``straight through'' intersection and an intersection with a
``collision''. Both elements are needed, since they appear in the action
of the Hamiltonian constraint.

\begin{figure}
${}^{}$\hspace{4cm}\epsfxsize=170pt \epsfbox{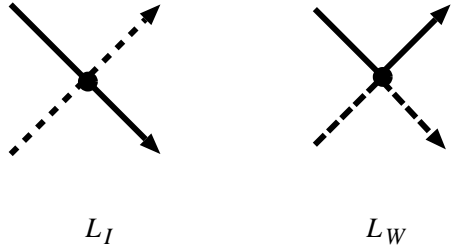}
\caption{The elements $L_{I}$ and $L_W$ that intervene in the
skein relation of the Conway polynomial with intersections.}
\end{figure}

The skein relations introduced above for the polynomial
imply the following relations for $a_2$, the second coefficient,

\begin{eqnarray}
a_2(L_+)-a_2(L_-) &=& a_1(L_0)\label{subibaja}\\
a_2(L_I)&=&{1\over 2} (a_2(L_+)+a_2(L_-)) \label{intersect}\\
a_2(L_W)&=&a_2(L_0)\label{collision}\\
a_2({\rm unknot})&=&0
\end{eqnarray}

Contrary to the relations for the full polynomial, the ones for the
second coefficient {\em are} uniquely determined by perturbative
calculations, up to irrelevant factors involving the number of
connected components of the loop. The first relation relates $a_2$
with $a_1$, the first coefficient, evaluated on an $L_0$.  When one
replaces a crossing in a knot diagram by an $L_0$ one generically is
left with a link composed by two loops.  In that case $a_1$ is
identical to the linking number of the two loops in the link. If the
resulting knot has a single component $a_1$ is zero.  The linking
number of two loops $lk(\gamma_1,\gamma_2)$ has the property of being
``Abelian'' $lk(\gamma_1\circ\gamma_2,\gamma_3)=
lk(\gamma_1,\gamma_3)+lk(\gamma_2,\gamma_3)$.  Also, for rerouted
loops, $lk(\gamma^{-1},\eta)=-lk(\gamma,\eta)$. All these properties
will be crucial for the calculations that follow.

Notice also that relations (\ref{subibaja},\ref{intersect}) imply that
one can replace an intersection by an upper or under crossing at the
cost of introducing terms involving linking numbers,

\begin{equation}
a_2(L_I)=a_2(L_-)+{1\over 2} a_1(L_0)=a_2(L_+)-{1\over 2} a_1(L_0)\label{+-}
\end{equation}

\begin{figure}[b]\epsfxsize=400pt \epsfbox{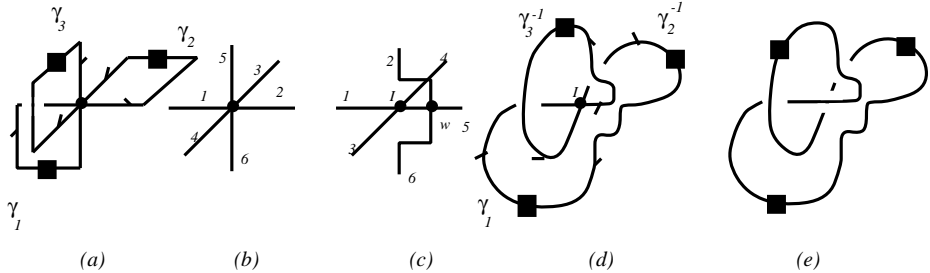}
\vspace{0.1cm}
\caption{The action of the Hamiltonian. Given a generic loop with a
triple intersection (a), the Hamiltonian
splits the triple intersection (b) into two double ones (c). For the case
of the second coefficient of the Conway polynomial the resulting
loop can be rearranged into a loop without intersections using the
skein relations. First one uses the skein relation for an intersection
with a kink and obtains loop (d).  Then one uses the skein relation for
regular intersections to convert the loop to a loop without intersections.
 The result is the second coefficient
 evaluated on a loop without
intersections plus contributions of linking numbers. The black squares
indicate that the different lobes of the loop could have arbitrary knottings
and interlinkings. The result only depends
on the local connectivity of the intersection.}
\label{a2fig}
\end{figure}

Let us apply the Hamiltonian constraint to the $a_2$ at a point with a
triple ``straight through'' intersection.  The loop in the lattice is
shown in figure \ref{a2fig}a and the detail of the intersection is
shown in figure \ref{a2fig}b, the numbers indicating the orientation
of the loop. We will only show in detail one half of one of the
contributions, that corresponding to when the Hamiltonian deforms ``to
the right'' in the ``1256'' plane in the notation of figure
\ref{a2fig}b. The action of the constraint produces a change in the
intersection and a reorientation shown in figure \ref{a2fig}c.  We
will now replace the intersection marked as $w$ in the figure, where a
collision takes place, with the corresponding skein relation
(\ref{collision}). The topology of the resulting loop is shown in
figure
\ref{a2fig}d (we draw it as a smooth loop in the continuum just for easing the
visualization process). In the resulting knot, which now has a single
intersection, denoted as $I$ in the figure, we replace it using
(\ref{+-}). This produces two kinds of terms. One is $a_2$ evaluated on
a loop without any intersection. This term cancels with the corresponding
contribution when the Hamiltonian deforms ``to the left'' in the same plane.
The other terms produced are linking numbers of the two subloops determined
by the intersection $I$.
This  contribution is given by to $-lk(\gamma_3^{-1},
\gamma_2^{-1}\circ\gamma_1)$ as can be checked by inspection of
figure \ref{a2fig}. Using the ``Abelian'' properties of the linking number
we discussed above, the
resulting contribution can be rewritten as
$-lk(\gamma_1,\gamma_3)+ lk(\gamma_2,\gamma_3)$. The action of the
Hamiltonian ``to the left'' in the same ``1256'' plane yields, after a
similar computation,
$-lk(\gamma_2^{-1},
\gamma_1\circ\gamma_3^{-1})=-lk(\gamma_2,\gamma_3)+lk(\gamma_1,\gamma_2)$.
The total contribution of the action of the Hamiltonian along the plane
``1256''
is therefore $-lk(\gamma_1,\gamma_3)+lk(\gamma_1,\gamma_2)$.

A similar calculation can be straightforwardly performed for the
contributions stemming from the action along the other two other
planes at the intersection. The result is that all the contributions
cancel each other.  This completes the proof.

In spite of the great differences with the formal calculation in the
continuum, there are some  remarkable similarities.
In the continuum, in terms
of either extended loop coordinates or multitangents
\cite{BrGaPuprl,DiGaGr} the second coefficient of the Conway polynomial
consisted in two terms one involving multitangents of order four and
one of order three. When the Hamiltonian acted it produced terms of
order three, four and five. The terms of order three and four
cancelled among themselves given the resulting topologies of the
involved loops (as is the case here for the contributions proportional
to $a_2$). The terms of rank five in the multitangents combined into a
series of linking numbers that cancelled among themselves due to the
Abelian nature of the linking numbers. This is exactly what we are
seeing here in the lattice calculation.

The lattice calculation is simple enough as to consider what would happen
for the next coefficient, the $a_3$. In that case, one could also
arrange several cancellations given the topologies of the loops, but then
one would be left not only with linking numbers but also with combinations
of $a_2$'s. Because the $a_2$ does not share the Abelian character with the
linking number it is unlikely that the resulting terms will cancel. A
related formal calculation in terms of extended loops shows precisely that
behavior \cite{Gr}.

We have performed the calculation for a ``straight through''
intersection.  What happens if the original intersection has ``kinks''
or  ``collisions''? A detailed calculation shows the action of the
Hamiltonian also vanishes. Crucial for this calculation is to use the
appropriate skein relations for intersections with kinks, as derived
{}from the expectation value of a Wilson loop in Chern-Simons theory. It
turns out that these relations are quite nontrivial and different from
the case of intersections that are ``straight through''. In particular,
they distinguish intersections in which the kinks are coplanar from
those in which they are not. This kind of result is only manifest in the
lattice and suggests that ``skein relations'' for intersecting knots
involve more than simply considering the planar projection of a knot, as
happened in the case without intersections.

The main objection to the result we have found is that it is tied to a
particular Hamiltonian. This Hamiltonian is constructed using the same
ideas that led to a lattice framework in which the diffeomorphism
constraint has desirable properties \cite{FoGaPu} and therefore might
suggest it is the appropriate one to deal with quantum gravity. It is
yet to be checked, however, if the constraint proposed closes the
appropriate algebra of constraints (at least in the continuum limit).
However, this is a problem that up to the moment has escaped solution in
all frameworks in which solutions to the constraints have been found.
Yet, it is a decidable matter in the framework presented hee.

An important lesson to be learnt from the use of the lattice
techniques is the much richer structure that skein relations for
intersecting knots have with respect to usual non intersecting
ones. The intersecting knot invariants considered here were specially
tailored to be compatible with the Mandelstam identities of
wavefunctions in the loop representation since they were obtained via
the loop transform in a perturbative framework. For coefficients
higher than $a_2$ it is clear that a more rigorous approach is needed
and lattice calculations might be a viable scenario for investigating
these issues.

We therefore see that the lattice formulation of the loop representation is
a powerful tool, where one can prove in a natural, simple and yet rigorous
way, formal results of the continuum. It leads also to insights into the
details of the construction of intersecting knot theory, which is the
kinematical arena of quantum gravity in the loop representation after
diffeomorphism invariant has been imposed.

This work was supported in part by grants NSF-INT-9406269,
NSF-PHY-9423950, NSF-PHY-9396246, research funds of the Pennsylvania
State University, the Eberly Family research fund at PSU and PSU's
Office for Minority Faculty development. JP acknowledges support of
the Alfred P. Sloan foundation through a fellowship. We acknowledge
support of Conicyt and PEDECIBA (Uruguay).

\end{document}